\documentstyle[12pt]{article} 
\def\beq{\begin{equation}}
\def\eeq{\end{equation}}

\def\al{\alpha}
\def\bt{\beta}

\def\De{\Delta}

\def\te{\theta}

\def\lam{\lambda}

\def\ep{\epsilon}

\def\l{\left (}
\def\r{\right )}
\def\fr{\frac}
\def\la{\label}
\def\hs{\hspace}
\def\vs{\vspace}

\def\ran{\rangle}
\def\lan{\langle}
\def\ov{\overline}
\def\tl{\tilde}
\def\tm{\times}

\begin{document}

\begin{titlepage}

\begin{center}
{\Large\bf    
Anomalous ${\cal U}(1)$: Solving Various Puzzles\\
\vspace{0.2cm}
Of MSSM And $SU(5)$ GUT}\footnote{To appear in the proceedings of NOON2001
Workshop held in Kashiwa, Japan, 5-8 Dec. 2001. }
\end{center}
\vspace{0.5cm}
\begin{center}
{\large 
{}~Qaisar Shafi$^{a}$
\footnote {E-mail address: shafi@bartol.udel.edu},~ 
{}~Zurab Tavartkiladze$^{b, c}$
\footnote {E-mail address: Z.Tavartkiladze@ThPhys.Uni-Heidelberg.DE} 
}
\vspace{0.5cm}

$^a${\em Bartol Research Institute, University of Delaware,
Newark, DE 19716, USA \\

$^b$ Institute for Theoretical Physics, Heidelberg University, 
Philosophenweg 16, \\
D-69120 Heidelberg, Germany\\

$^c$ Institute of Physics,
Georgian Academy of Sciences, Tbilisi 380077, Georgia\\
}

\end{center}
\vspace{1.0cm}

\begin{abstract}

We discuss how an anomalous ${\cal U}(1)$ symmetry when appended to MSSM
and SUSY GUTs [e.g. $SU(5)$] can help overcome a variety of puzzles
related to charged fermion masses and mixings, flavor changing processes,
proton decay and neutrino oscillations. Proton lifetime for $SU(5)$ GUT,
in particular, is predicted in a range accessible to the ongoing or
planned searches.

\end{abstract}

\end{titlepage}

\section{Introduction: Some Puzzles of MSSM and Beyond}

The standard model provides an excellent description of almost all
present
experimental data. Supersymmetry
(SUSY) is highly motivated because it offers possibility of resolving
the gauge hierarchy problem. Furthermore, superstring theories are
believed to be good candidates for a unified description of gauge theories
and quantum gravity. Therefore, for realistic model building including 
SUSY is
a good way to proceed and SUSY GUTs provide an excellent example.
However, supersymmetry introduces several problems and puzzles which
require explanations. The puzzles can be divided in two categories:
{\bf 1} puzzles which are common only to SUSY theories, i.e. those which
appear due to SUSY extension (SUSY puzzles); and {\bf 2} The puzzles
which exist also
in non-SUSY theories and within SUSY extensions have the same status
(non-SUSY puzzles). From
puzzles {\bf 1}, {\bf 2} we list here only those which we will attempt
to resolve.  

{\bf 1~SUSY puzzles}

~~~a)~Problem of FCNC arises because with SUSY there is a new source for
flavor violation. Apriori, without any specific arrangement there is no
universality and an alignment (which would guarantee flavor
conservation) in the soft SUSY breaking terms. This can create 
non-diagonal
gaugino-fermion-sfermion interactions, leading to the FCNC \cite{mas}.

~~~b)~$d=5$ Baryon number violation;

With SUSY there is a new source for baryon number violation. Namely $d=5$
operators \cite{wein}
\beq
\fr{\lam }{M}qqql~,~~~~~~~\fr{\lam'}{M}u^cu^cd^ce^c~,
\la{d5bar}
\eeq
where $M$ is some cutoff mass scale and $\lam $, $\lam'$ are dimensionless
couplings depending on the model. In GUTs,  $M$ usually stands for
colored triplet masses. In minimal $SU(5)$ the nucleon decay mainly
proceeds
through the channel $p\to K\nu $ and its lifetime is estimated to be
$10^{29\pm 2}$~yr \cite{time}, \cite{d5flavor}, which is embarrassingly
small in comparison
with the latest SuperKamiokande (SK) limit
$\tau_p^{\rm exp}\stackrel{>}{_\sim }10^{33}$ yr \cite{pdg}. Therefore,
some mechanism (with reasonable extension) 
\cite{d5flavor}, \cite{d5sup1}, \cite{8} must be applied
to suppress these contributions. Once colored triplet induced $d=5$
operators are properly suppressed, one has to take care of Planck scale
$d=5$ baryon number violating operators. In (\ref{d5bar}) with
$M=M_{Pl}=2.4 \cdot 10^{18}$~GeV, in order to satisfy the experimental
bounds
we need $\lam , \lam'\stackrel{<}{_\sim }10^{-8}$. This kind of
suppression requires additional explanation 
\cite{d5flavor}, \cite{gaugeZ}-\cite{1}.

{\bf 2 ~Non-SUSY puzzles} 

~~~a)~Problem of flavor - hierarchies between charged fermion masses and
mixings;

In the charged fermion sector there are noticeable hierarchies within 
the 
fermion Yukawa couplings and the CKM matrix elements. Since the mass 
of the top quark is close to the electroweak symmetry breaking scale
($\sim 100$~GeV), its Yukawa coupling is of order unity ($\lam_t \sim 1$).
As far as the Yukawa couplings of the $b$ quark and $\tau $ lepton are
concerned, their values could vary in a range 
$\lam_b \sim \lam_{\tau }\sim 10^{-2}-1 $, depending on the value of 
the MSSM parameter $\tan \bt $ ($\sim 1 - 60$). Introducing the
dimensionless parameter
$\ep \simeq 0.2$, 
one can express the observed hierarchies between the charged fermion
Yukawa couplings as follows:
\beq
\lambda_t\sim 1~,~~
\lambda_u :\lambda_c :\lambda_t \sim
\epsilon^6:\epsilon^3 :1~,
\la{ulam}
\eeq
\beq
\lam_b\sim \lam_{\tau}\sim \lam_t\fr{m_b}{m_t}\tan \bt ~,~~~
\lambda_d :\lambda_s :\lambda_b \sim
\epsilon^5:\epsilon^2 :1~,
\label{dlam}
\eeq
\beq
\lambda_e :\lambda_{\mu } :\lambda_{\tau } \sim
\epsilon^5:\epsilon^2 :1~,
\label{elam}
\eeq
while for the CKM matrix elements:
\beq
V_{us}\sim \epsilon~,~~~V_{cb}\sim \epsilon^2~,~~~
V_{ub}\sim \epsilon^3~.
\label{ckm}
\eeq
In constructing models, one should arrange for a natural understanding of
the hierarchies in (\ref{ulam})-(\ref{ckm}).

~~~b)~Atmospheric and Solar Neutrino puzzles;

The latest atmospheric and solar neutrino data (see \cite{atm} and \cite{sol}
respectively) seem to provide convincing evidence for the phenomena of
neutrino
oscillations. Ignoring the LSND data, the atmospheric and solar neutrino
anomalies can be explained within the three states of active neutrinos.
In this paper we will study oscillation scenarios
without the sterile neutrinos, which are disfavored by the
data \cite{sol, atm}. 

The atmospheric neutrino data suggest oscillations of $\nu_{\mu }$ into $\nu_{\tau }$,
with the following oscillation parameters:
$$
{\cal A}(\nu_{\mu }\to \nu_\tau )\equiv \sin^2 2\te_{\mu \tau}
\simeq 1~,
$$
\beq
\De m^2_{\rm atm}\simeq 3\cdot 10^{-3}~{\rm eV}^2~.
\la{atmdat}
\eeq

The solar neutrino anomaly seems consistent with oscillation scenarios, amongst which the
most likely seems to be the large angle MSW (LMA) oscillation of $\nu_e$
into $\nu_{\mu , \tau }$ \cite{sol}, with the oscillation parameters:
$$
{\cal A}(\nu_e\to \nu_{\mu, \tau } )\equiv \sin^2 2\te_{e \mu , \tau}
\approx 0.8~, 
$$
\beq
\De m^2_{\hs{0.5mm}\rm sol}\sim 10^{-4}~{\rm eV}^2~.
\la{LAMSWdat}
\eeq
The scenario of low MSW (LOW) oscillations of solar neutrinos
require:
$$  
\sin^2 2\te_{e \mu , \tau}
\simeq 1.0~,
$$  
\beq
\De m^2_{\hs{0.5mm}\rm sol}\simeq 8\cdot 10^{-8}~{\rm eV}^2~.
\la{LOWMSWdat}
\eeq
Let us note that the small angle MSW and large angle vacuum oscillation
solutions seem to be disfavored by data.

It is worth noting that within MSSM, the neutrinos acquire masses only through
non-renormalizable $d=5$ Planck scale operators $l_il_jh_u^2/M_P$ which,
for
$\lan h_u^0\ran \sim 100$~GeV and $M_P=2.4\cdot 10^{18}$~GeV (reduced
Planck
mass)
give $m_{\nu_i }\sim 10^{-5}$~eV. Therefore,  for  (\ref{atmdat})  
or (\ref{LAMSWdat}) we need physics beyond the MSSM.
In order to generate the appropriate neutrino masses, we will introduce
heavy right handed neutrino states ${\cal N}_i$. The `light' 
left-handed 
neutrinos will acquire masses through the see-saw mechanism
\cite{seesaw}.

In building neutrino oscillation scenarios, the main challenge is to generate
desirable magnitudes for neutrino masses and their mixings, and to
understand why in some cases the mixing angles are large (and even
maximal), while the quark CKM matrix elements (\ref{ckm}) are suppressed.
Below we will present a mechanism which successfully resolves all
of these problems.

~~~c)~Wrong asymptotic relations originating from GUTs.

Within GUTs the quark-lepton families are embedded in unified multiplets
and
because of this, wrong asymptotic relations for masses and mixings 
are possible unless some care is exercised. In minimal $SU(5)$ chiral
matter
consists of anomaly free $10+\bar 5$ supermultiplets per generation, where
$10=(q, u^c, e^c)$, $\bar 5 =(l, d^c)$. The couplings generating the
up, down quark and charged lepton masses are respectively
$10\cdot 10 \cdot H+10\cdot\bar 5\cdot \bar H$, ($H$, $\bar H $ are
'higgs' superfields 
in $5$ and $\bar 5$ representations). The second coupling 
gives
$\hat{M}_f^0=\hat{M}_d^0$ at GUT scale, which for the third generation
yields the
reasonable asymptotic relation $m_b^0=m_{\tau }^0$, but for light
generations it gives $\l\fr{m_d}{m_s}\r^0 =\l\fr{m_e}{m_{\mu }}\r^0$ which
is
unacceptable. For improving this picture,  either a 
scalar $45$ plet \cite{45pl}, or  vector-like matter \cite{9},
\cite{8}, 
or some
non-renormalizable operators \cite{nonren}
can be employed.


As we see, for solving the problems listed above, an extension 
of the minimal scheme is
needed. Otherwise, in some cases, we should simply assume
the presence of appropriately suppressed couplings and mass scales [for
instance for 
{\bf 1}.b)$, {\bf 2}.a), b)$]. The latter puzzle is tied to the so-called
naturalness issue \cite{natis}, namely why are some
couplings and scales small, when apriori there is no reason to expect
it? Below we discuss some extensions which provide natural mechanisms
for resolving the above-listed problems.

\section{Anomalous ${\cal U}(1)$ as a Flavor Symmetry and Mediator of
SUSY Breaking}

We introduce a ${\cal U}(1)$ gauge symmetry which acts as a
flavor symmetry and provides for a natural understanding of the
hierarchies
between charged
fermion masses and mixings. ${\cal U}(1)$ also turns out to be crucial for
building textures of neutrino mass matrices that provide scenarios for
simultaneous explanations of atmospheric and solar neutrino puzzles. It
will turn out that ${\cal U}(1)$ is anomalous, which allows us to use 
it as a mediator of SUSY breaking \cite{gia}, \cite{5}. Thanks to this,
the squarks and
sleptons which correspond to the light generations can have masses of
${\rm few}\cdot 10$~TeV (this value can be acceptable
also for $\tilde{b}^c$, $\tl{\nu }_{\tau }$ and $\tilde{\tau }$ in the low
$\tan \bt $
regime). Because of this, the FCNC can be adequately suppressed due to
the decoupling \cite{FCNC}-\cite{d5sup2}, \cite{5}. It turns
out
that together with this, some  nucleon decay modes are strongly
suppressed \cite{d5sup2}, \cite{5}. Namely, diagrams with heavy squarks
and/or sleptons inside the
loops decouple. As far as the Planck scale $d=5$ baryon number violating
operators are concerned, they can also be adequately suppressed by the 
${\cal U}(1)$ symmetry. Therefore, the advantages of ${\cal U}(1)$
symmetry are
manyfold, and we will
present an $SU(5)$ GUT to show how things work out.

The  anomalous
${\cal U}(1)$ factors can arise from string theories
\footnote{Recently, ref. \cite{1} presented an example
where anomalous ${\cal U}(1)$ arises in 4D level through 5D
orbifold compactification. The cancellation of anomalies
occur through bulk Chern-Simons term.}. Cancellation of the anomaly
occurs through the Green-Schwarz mechanism \cite{gs}. Due to the anomaly the
Fayet-Illiopoulos D-term
$\xi \int d^4\theta V_A$
is always generated, where in string theory $\xi $ is given by \cite{fi}

\begin{equation}
\xi =\frac{g_A^2M_P^2}{192\pi^2}{\rm Tr}Q~.
\label{xi}
\eeq
The $D_A$-term will have the form:

\begin{equation}
\frac{g_A^2}{8}D_A^2=\frac{g_A^2}{8}
\left(\Sigma Q_a|\varphi_a |^2+\xi \right)^2~,
\label{da}
\eeq
where $Q_a$ is the `anomalous' charge of $\varphi_a $ superfield.
For ${\cal U}(1)$ breaking we introduce the singlet superfield $X$
 with 
${\cal U}(1)$ charge $Q_X$. Assuming
$\xi >0~$ [${\rm Tr}Q>0$ in (\ref{xi})],
and taking
\beq
Q_X=-1~,
\la{take}
\eeq
the cancellation of $D_A$ in (\ref{da}) and nonzero $\lan X\ran$ are
ensured
($\langle X\rangle =\sqrt{\xi }$).
Further, we will take
\beq
\frac{\langle X\rangle }{M_P}\equiv \epsilon \simeq 0.2~,
\label{epsx}
\eeq
where $\ep $ turns out to be an important expansion parameter. 
Let us note that an anomalous ${\cal U}(1)$ for understanding the
hierarchies of fermion masses and mixings and a variety of neutrino
oscillation scenarios has been discussed in several papers 
of \cite{dem}-\cite{6},  \cite{8}, \cite{5}, \cite{9}, \cite{10}, \cite{11} 

\subsection{Neutrino oscillations and quark-lepton masses}

Let us begin with the neutrino sector and we first discuss
two ways of obtaining large/maximal neutrino mixings with the help
of ${\cal U}(1)$ flavor symmetry.
With two flavors of lepton doublets $l_1$ and $l_2$, one way of having large mixing is the so-called  {\it democratic approach}. 
Here the ${\cal U}(1)$ symmetry does not distinguish the two 
flavors \cite{dem}, i.e. they have the same ${\cal U}(1)$ charges
$Q_{l_1}=Q_{l_2}=n$(positive integer number). In this case, the expected
neutrino mass matrix will be:
\beq
\begin{array}{cc}
&  {\begin{array}{cc}
\hspace{-5mm}~~ & \,\,

\end{array}}\\ \vspace{2mm}
\begin{array}{c}
\\  \\
\end{array}\!\!\!\!\!\hat{m}_{\nu }=&{\left(\begin{array}{cc}
\,\,1~~ &\,\,1
\\
\,\,1~~ &\,\,1
\end{array}\right)\bar m\ep^{2n} }~,~~~\bar m=\fr{h_u^2}{\bar M}~,
\end{array}  \!\!
\la{demnu}
\eeq
with entries of order unity ($\bar M$ is some mass scale and
we have assumed $Q_{h_u}=0$).
Therefore, naturally large
$\nu_1-\nu_2$ mixing is expected, $\sin^2 2\te_{12}\sim 1$. 
Also, one can expect $m_{\nu_1}\sim m_{\nu_2}$, and if this mechanism is
used for atmospheric neutrinos, somehow one has to keep one state light, 
in order to accomodate also the solar neutrino puzzle. This can be done
\cite{1N, 9} by introducing a single right handed neutrino 
${\cal N}$.
After integrating it out, due to degeneracy only one state acquires a
non-zero mass. The remaining
states can be used for the solar neutrino puzzle. An appropriate mass scale for
the latter can be generated by introducing a relatively heavy right handed state
${\cal N}'$ with suppressed coupling to ${\cal N}$.

A different approach is the so-called 
{\it maximal mixing mechanism} \cite{maxmix}, \cite{10}, \cite{11}. It is
realized by
assigning different ${\cal U}(1)$ charges for the flavors 
$l_1, l_2$. Introducing two right handed states 
${\cal N}_1$, ${\cal N}_2$ and the following ${\cal U}(1)$ charge
prescriptions
$$
Q_{l_1}=k+n~,~~~Q_{l_2}=k~,~~~Q_{h_u}=0~,
$$
\beq
Q_{{\cal N}_1}=-Q_{{\cal N}_2}=k+k'~,
\la{maxQ}
\eeq
with $k, n, k'>0~, n\stackrel{>}{_{-}}k'$,
the `Dirac' and `Majorana' coupling are given by:
\beq
\begin{array}{cc}
 & {\begin{array}{cc}
{\cal N}_1~~~~&\,\,~~~{\cal N}_2 
\end{array}}\\ \vspace{2mm}
\begin{array}{c}
l_1\\ l_2 

\end{array}\!\!\!\!\! &{\left(\begin{array}{cc}
\,\, \epsilon^{2k+n+k'}~~ &
\,\, \epsilon^{n-k'}
\\
\,\, \epsilon^{2k+k'} ~~ &\,\,0
\end{array}\right)h_u }~,
\end{array}  \!\!~~~
\begin{array}{cc}
 & {\begin{array}{cc}
~{\cal N}_1~&\,\,
~~~{\cal N}_2~~~ 
\end{array}}\\ \vspace{2mm}
\begin{array}{c}
{\cal N}_1 \\ {\cal N}_2

\end{array}\!\!\!\!\! &{\left(\begin{array}{ccc}
\,\, \epsilon^{2(k+k')}
 &\,\,~~~1
\\
\,\, 1
&\,\,~~~0
\end{array}\right)M_{\cal N}~.
}
\end{array}
\la{maxcoupl}
\eeq
After integrating out the heavy ${\cal N}_1, {\cal N}_2$ states, the
neutrino mass matrix is given by
\beq
\begin{array}{cc}
&  {\begin{array}{cc}
\hspace{-5mm}~~ & \,\,

\end{array}}\\ \vspace{2mm}
\begin{array}{c}
\\  \\
\end{array}\!\!\!\!\!\hat{m}_{\nu }=&{\left(\begin{array}{cc}
\,\,\ep^n~~ &\,\,1
\\
\,\,1~~ &\,\,0
\end{array}\right)\bar m }~,~~~\bar m=\fr{h_u^2\ep^{2k+n}}{M_{\cal N}}
\end{array}  \!\!~~~~~,
\la{maxnu}
\eeq
a quasi off-diagonal form, leading to a mixing angle
\beq
\sin^2 2\te_{12}=1-{\cal O}(\ep^{2n})~,
\la{maxmix}
\eeq
which is close to maximal mixing. The form (\ref{maxnu}) is
guaranteed
by the appropriate zero entries in (\ref{maxcoupl}), which are
ensured by ${\cal U}(1)$ symmetry. This mechanism turns out to be
very convenient for achieving nearly maximal mixings between neutrino
flavors within various realistic models, such as $SU(5)$
\cite{9},
$SO(10)$ \cite{6}, $SU(4)_c\tm SU(2)_L\tm SU(2)_R$ 
\cite{11}, etc.

Returning to our scheme, we attempt to obtain the bi-maximal 
texture 
through ${\cal U}(1)$ flavor symmetry. For this, we will combine the two
mechanisms discussed above. Namely, the second and third lepton
doublet states will have the same ${\cal U}(1)$ charges, which will lead to their
large mixing. The state $l_1$ will have a suitable charge, one that ensures 
maximal $\nu_1 - \nu_2$ mixing.

Introducing two right handed ${\cal N}_{1, 2}$ neutrino states and
choosing ${\cal U}(1)$ charges as
$$
Q_X=-1~,~Q_{l_2}=Q_{l_3}=k~,~Q_{l_1}=k+n~,~Q_{h_u}=Q_{h_d}=0~,
$$
\beq
Q_{{\cal N}_1}=-Q_{{\cal N}_2}=k+k'~,
\label{charges}
\eeq
with
\beq
k, n, k'>0~,~~~~n\stackrel{>}{_{-}}k'~,
\la{condnk}
\eeq
the `Dirac' and `Majorana' couplings will have forms:

\begin{equation}
\begin{array}{cc}
 & {\begin{array}{cc}
{\cal N}_1~~&\,\,~~~{\cal N}_2~~
\end{array}}\\ \vspace{2mm}
\begin{array}{c}
l_1\\ l_2 \\ l_3

\end{array}\!\!\!\!\! &{\left(\begin{array}{ccc}
\,\, \epsilon^{2k+n+k'}~~ &
\,\, \epsilon^{n-k'}
\\
\,\, \epsilon^{2k+k'} ~~ &\,\,0
\\
\,\, \epsilon^{2k+k'} ~~ &\,\,0
\end{array}\right)h_u }~,
\end{array}  \!\!~~~
\begin{array}{cc}
 & {\begin{array}{cc}
{\cal N}_1~&\,\,
~~~{\cal N}_2~~~
\end{array}}\\ \vspace{2mm}
\begin{array}{c}
{\cal N}_1 \\ {\cal N}_2

\end{array}\!\!\!\!\! &{\left(\begin{array}{ccc}
\,\, \epsilon^{2(k+k')}
 &\,\,~~~1
\\
\,\, 1
&\,\,~~~0
\end{array}\right)M_{\cal N}~
}
\end{array}~.
\label{Ns}
\eeq
After integrating out ${\cal N}_{1, 2}$, we obtain the texture

\begin{equation}
\begin{array}{ccc}
&  {\begin{array}{ccc}
\hspace{-5mm}~~ & \,\, & \,\,

\end{array}}\\ \vspace{2mm}
\begin{array}{c}
 \\  \\
 \end{array}\!\!\!\!\!\hat{M}_{\nu }\propto &{\left(\begin{array}{ccc}
\,\,\epsilon^n~~ &\,\,1~~ &
\,\,1
\\
\,\,1~~   &\,\,0~~  &
\,\,0
 \\
\,\,1~~ &\,\,0~~ &\,\,0
\end{array}\right) }m~,~~~~~~m=\frac{\epsilon^{2k+n}h_u^2}{M_{\cal N}}~,
\end{array}  \!\!
\label{u1nu}
\eeq
In (\ref{u1nu}) coefficients of order
unity are assumed. Without (1, 1) entry the (\ref{u1nu}) has 
$L_e-L_{\mu }-L_{\tau }$ symmetry, which also can be used \cite{Lemu}
for obtaining bi-maximal texture. In our case deviation from (1, 1)
zero entry is controlled by ${\cal U}(1)$ flavor symmetry \cite{3}.
The nonzero (1, 1) entry in (\ref{u1nu}) 
guarantees that $\De m_{12}^2 {\neq}0$. 
Using (\ref{u1nu}) the oscillation parameters are:
$$
\Delta m^2_{32}\equiv m_{\rm atm}^2= m^2\sim 10^{-3}~{\rm eV}^2~,
$$
\beq
{\cal A}(\nu_{\mu }\to \nu_{\tau })\sim 1~,
\label{atmosc}
\eeq
$$
\Delta m^2_{21 }\simeq 2m_{\rm atm}^2\epsilon^n~,
$$
\beq
{\cal A}(\nu_e \to \nu_{\mu , \tau }) =1-{\cal O}(\epsilon^{2n})~.
\label{solosc}
\eeq
Note that the model does not constrain $n$ for the time being. So, 
LMA and LOW solutions for solar neutrinos can be realized.
With prescription (\ref{charges}), the expected contribution from the
charged
lepton sector to the angles $\te_{23}^l$ and $\te_{12}^l$ will be
$\sim 1$ and $\sim \ep^n$ respectively. These do not alter 
the expressions in (\ref{atmosc}), (\ref{solosc}).

The ${\cal U}(1)$ charge selection in (\ref{charges}) nicely blends with
the charged
fermion sector. Indeed, considering the following prescription:
$$
Q_{q_3}=0~,~~Q_{q_2}=2~,~~Q_{q_1}=3~,~~
Q_{d^c_3}=Q_{d^c_2}=p+k~,~~
$$
$$
Q_{d^c_1}=p+k+2~,~~Q_{u^c_3}=0~,~~Q_{u^c_2}=1~,~~Q_{u^c_1}=3~,
$$
\beq
Q_{e_3^c}=p~,~~~Q_{e_2^c}=p+2~,~~~
Q_{e_1^c}=p+5-n~,~~~
\label{qch}
\eeq
the structures of Yukawa matrices for up-down quarks and charged leptons
are respectively:

\begin{equation}
\begin{array}{ccc}
 & {\begin{array}{ccc}
\hspace{-5mm} u^c_1 & \,\,~u^c_2 ~~ & \,\,u^c_3 ~

\end{array}}\\ \vspace{2mm}
\begin{array}{c}
q_1 \\ q_2 \\q_3
 \end{array}\!\!\!\!\! &{\left(\begin{array}{ccc}
\,\,\epsilon^6~~ &\,\,\epsilon^4~~ &
\,\,\epsilon^3  
\\  
\,\,\epsilon^5~~   &\,\,\epsilon^3~~  &
\,\,\ep^2
 \\
\,\,\epsilon^3~~ &\,\,\epsilon ~~ &\,\,1
\end{array}\right)h_u }~, 
\end{array}  \!\!  ~~~~~
\label{up}
\eeq

\begin{equation}
\begin{array}{ccc}
 & {\begin{array}{ccc}
\hspace{-5mm} d^c_1~ & \,\,d^c_2 ~~ & \,\,d^c_3 ~~~~~~

\end{array}}\\ \vspace{2mm}
\begin{array}{c}  
q_1 \\ q_2 \\q_3
 \end{array}\!\!\!\!\! &{\left(\begin{array}{ccc} 
\,\,\epsilon^5~~ &\,\,\epsilon^3~~ &
\,\,\epsilon^3
\\
\,\,\epsilon^4~~   &\,\,\epsilon^2~~  &
\,\,\epsilon^2
 \\
\,\,\epsilon^2~~ &\,\,1~~ &\,\,1
\end{array}\right)\epsilon^{p+k}h_d }~,
\end{array}  \!\!  ~~~~~
\label{down}
\eeq

\begin{equation}
\begin{array}{ccc}
 & {\begin{array}{ccc}
\hspace{-7mm} e^c_1~~~~~ & \,\,e^c_2 ~~~ & \,\,e^c_3 ~~
  
\end{array}}\\ \vspace{2mm}
\begin{array}{c}
l_1 \\ l_2 \\l_3
 \end{array}\!\!\!\!\! &{\left(\begin{array}{ccc}
\,\,\epsilon^5~~ &\,\,\epsilon^{n+2}~~ &
\,\,\epsilon^n
\\  
\,\,\epsilon^{5-n}~~   &\,\,\epsilon^2~~  &
\,\,1
 \\
\,\,\epsilon^{5-n}~~ &\,\,\epsilon^2~~ &\,\,1
\end{array}\right)\epsilon^{p+k}h_d }~.
\end{array}  \!\!  ~~~~~
\label{lept}
\eeq
Upon diagonalization of (\ref{up})-(\ref{lept}) it is easy to verify that
the desired relations (\ref{ulam})-(\ref{ckm}) for the Yukawa couplings and CKM
matrix elements are realized. From (\ref{down}), (\ref{lept}) we have
\beq
\tan \bt \sim \ep^{p+k}\fr{m_t}{m_b}~.
\label{tangens}
\eeq
As we previously mentioned, MSSM does not fix the values of $n, k, p$ in
(\ref{charges}), (\ref{qch}). Because of this, the solar neutrino oscillation
scenario is not
specified and  both
LMA and LOW are possible solutions. 

\subsection{$D_A$-term SUSY Breaking
Suppression of FCNC and Nucleon Decay}

The cancellation of $D_A$-term (\ref{da}) was ensured by the VEV of $X$
(at this stage we do not consider any
superpotential couplings for $X$). Including a mass term for $X$
in the superpotential,
\beq
W_m=\frac{m}{2}X^2~,
\label{massx}
\eeq
the cancellation of $D_A$ will be partial, and SUSY will be broken
due to non-zero $F$ and $D$ terms. Taking into account 
(\ref{da}) and (\ref{massx}), we have
\beq
X^2=\xi -\frac{4m^2}{g_A^2}~,~~~
\langle D_A\rangle =\frac{4m^2}{g_A^2}~,~~~~
\langle F_X\rangle \simeq m\sqrt{\xi }~.
\label{solx}
\eeq
{}From (\ref{da}), taking into account (\ref{solx}), 
for the soft scalar masses squared (${\rm mass}^2$) we have
\beq
m^2_{\tilde{\varphi }_a}=Q_am^2~.
\label{masssc}
\eeq
Thus, the scalar components of superfields which have non-zero
${\cal U}(1)$ charges gain masses through $\langle D_A\rangle$.
We will assume that the scale $m$ is in the range $\sim 10$~TeV. 
Those states
which have zero ${\cal U}(1)$ charges will gain soft masses
of the order of gravitino mass $m_{3/2}$ from the K\"ahler potential 
\beq
m_{3/2}=\frac{F_X}{\sqrt{3}M_P}=m\frac{\epsilon }{\sqrt{3}}~,
\label{gravmass}
\eeq
which, for $m=10$~TeV, is relatively
suppressed ($\sim 1$~TeV). 
The gaugino masses also will have the same magnitudes
\beq
M_{\tilde{G}_i}\sim m_{3/2}\sim 1~{\rm TeV}~.
\label{gaugmass}
\eeq

The mass term (\ref{massx}) violates the ${\cal U}(1)$ symmetry
and is taken to be in the $10$~TeV range. Its origin 
may lie in a strong
dynamics where $m$ is replaced by the VEV of some superfield(s). 
We do not present here any examples of this and refer the reader to
\cite{gia}, \cite{5} for detailed discussions.
The important point is that the soft masses (\ref{masssc}) of sparticles
are
controlled by the ${\cal U}(1)$ symmetry. 

Turning now to the question of FCNC, we require that the
`light' quark-lepton 
superfields carry non-zero 
${\cal U}(1)$ charges. This means that the soft masses of their scalar 
components are in the $10$~TeV range, which automatically 
suppresses  flavor 
changing processes  such as $K^0-\bar K^0$, $\mu \to e\gamma$
etc., thereby satisfying the present experimental bounds \cite{mas}.
To prevent upsetting the gauge hierarchy, the third generation
up squarks must have masses no larger than a TeV or so \cite{dimgiu}
(hence they have zero ${\cal U}(1)$ charge). The same applies to  
sbottom and stau for large $\tan \beta $
since, for $\lambda_b\sim \lambda_{\tau }\sim 1$, 
large masses ($\stackrel{>}{_\sim }10$~TeV) of $\tilde{b}$
and $\tilde{\tau }$ would spoil the gauge hierarchy.  

Although the tree level mass of the stop can be arranged to be in the 
$1$~TeV range by the ${\cal U}(1)$ symmetry, the $2$-loop contributions
from heavy sparticles of the first two generations can 
drive the stop ${\rm mass}^2$  
negative \cite{dimgiu}. This is clearly unacceptable, and one proposal 
for avoiding it \cite{hisano}
requires the existence of new states in the multi-TeV range.
This type of extension makes the decoupled solution viable and
realistic models can be built \cite{5}.

Let us now turn to some implications for proton decay.
We assume that $d=5$ baryon number 
violating operators arise from the couplings
\beq
qAqT+qBl\bar T~,
\label{qqt}
\eeq
after integration of color triplets $T, \bar T$ with mass 
$M_T\sim 2\cdot 10^{16}$~GeV (we first consider triplet 
couplings with left-handed matter).
After wino dressing of appropriate $d=5$ operators, the resulting $d=6$ 
operators causing proton to decay into a 
neutrino and charged lepton channels have the respective forms
\cite{time}, \cite{d5flavor}:
\beq
\frac{g_2^2}{M_T}
\alpha (u_a d^i_b)(d^j_c\nu^k) 
\varepsilon^{abc}~,
\label{d6nu}
\eeq
 \beq
\frac{g_2^2}{M_T}
\alpha' (u_a d^i_b)
(u_ce^j)
\varepsilon^{abc}~,
\label{d6e}
\eeq
where
$$
\alpha=
-[(L_d^{+}\hat{B}L_e)_{jk}
(L_u^{+}\hat{A}L_d^{*})_{mn}+
(L_d^{+}\hat{A}L_u^{*})_{jm}
(L_d^{+}\hat{B}L_e)_{nk}]
V_{mi}(V^{+})_{n1}
I(\tilde{u}^m,\tilde{d}^n)+ 
$$
\beq
[(L_u^{+}\hat{A}L_d^{*})_{1i}
(L_u^+\hat{B}L_e)_{mk}
-(L_d^{+}\hat{A}L_u^{*})_{im}
(L_u^{+}\hat{B}L_e)_{ik}]V_{mj}
I(\tilde{u^m},\tilde{e^k})~, 
\label{d6nu1}
\eeq
$$
\alpha' =
[-(L_u^{+}\hat{A}L_d^{*})_{1i}
(L_d^{+}\hat{B}L_e)_{mj}
+(L_u^{+}\hat{A}L_d^{*})_{1m}
(L_d^{+}\hat{B}L_e)_{ij}](V^{+})_{m1}
I(\tilde{d^m},\tilde{\nu^j})+ 
$$
\beq
[(L_u^{+}\hat{B}L_e)_{1j}
(L_u^{+}\hat{A}L_d^{*})_{mn}+
(L_u^{+}\hat{A}L_d^{*})_{1m}
(L_e^{T}\hat{B}^TL_u^{*})_{jn}]
(V^{+})_{m1}
V_{ni}
I(\tilde{d}^m, \tilde{u}^n)~. 
\label{d6e1}
\eeq
$L$'s are unitary matrices which rotate the left handed fermion states
to diagonalize the mass matrices, and $I$'s 
are functions obtained after loop integration and depend on the SUSY 
particle masses circulating inside the loop. For example \cite{time},
\beq
I(\tilde{u}, \tilde{d})=\frac {1}{16\pi^2}\frac
{m_{\tilde{W}}}{m_{\tilde{u}}^2- m_{\tilde{d}}^2} 
\left ( \frac{m_{\tilde{u}}^2}{m_{\tilde{u}}^2- 
m_{\tilde{W}}^2}\ln \frac{m_{\tilde{u}}^2}{m_{\tilde{W}}^2}- 
\frac{m_{\tilde{d}}^2}{m_{\tilde{d}}^2- 
m_{\tilde{W}}^2}\ln \frac{m_{\tilde{d}}^2}{m_{\tilde{W}}^2} \right )~,
\label{int}
\eeq
with similar expressions for $I(\tilde{d}, \tilde{\nu} )$ and
$I(\tilde{u},\tilde{e} )$.

Consider those  
diagrams in which  sparticles of the first two 
families participate. Since their masses are large 
($\stackrel{>}{_\sim }10$~TeV) compared 
to the case with minimal $N=1$ SUGRA, we expect considerable  
suppression of proton decay.  
For minimal $N=1$ SUGRA, 
$m_{\tilde{u}}\sim m_{\tilde{d}}\sim m_{\tilde{W}}\sim m_{3/2}\sim 1$~TeV,
and (\ref{int}) can be approximated by
\beq
I_0\approx \frac{1}{16\pi^2}\frac{1}{m_{3/2}}~.
\label{int0}
\eeq 
In the ${\cal U}(1)$ mediated SUSY breaking scenario, 
expression  (\ref{int}) takes the  
form
\beq
I'\approx \frac{1}{16\pi^2}\frac{m_{\tilde{W}}}{m_{\tilde{q}}^2}
\equiv \eta I_0
\label{int1}
\eeq
The nucleon lifetime in this case will be enhanced by the factor
$\frac{1}{\eta^2}\sim 10^4$.

Of course, there exist diagrams in which one sparticle from the 
third and one from the `light' families participate. In this case,
(\ref{int}) takes the form
\beq
I''\approx \frac{1}{16\pi^2}\frac{2m_{\tilde{W}}}{m_{\tilde{q}}^2}
\ln \frac{m_{\tilde{q}}}{m_{\tilde{W}}}
\equiv \eta' I_0
\label{int2}
\eeq
and the corresponding proton lifetime will be
$\sim \frac{1}{\eta'^2}\sim 500$ times larger. 
This suppression is enough to bring the proton lifetime near the
experimental
limit, providing an opportunity for testing this type of scenario in the
near future.

As pointed out in \cite{time}, \cite{d5flavor} (within minimal $N=1$
SUGRA), 
the contribution 
from diagrams in which sparticles from the third 
generation participate could be comparable with those 
arising from the light 
generation sparticle exchange. 
Namely, second term of (\ref{d6nu1}) with $\tilde{t}$, $\tl{\tau }$ inside
the loop give a contribution comparable to diagrams with first two
generation sparticles circulating inside the loop. Since minimal SUSY
$SU(5)$
gives unacceptably fast proton decay with $\tau_0\sim 10^{29\pm 2}$~yr, 
care must be exercised in realistic model building 
(the situation is exacerbated if $\tan \beta $ is large).
This problem is easily avoided in the anomalous ${\cal U}(1)$
mediated SUSY breaking scenario. Note that  
in second term of (\ref{d6nu1}) the $I$ depends on mass of $\tl{e}^k$
state and
even if the latter belongs to the third family, it can have mass in the
$10$~TeV 
range if $\tan \beta $
is either of intermediate ($\sim 10-15$) or low value
(this is required for preserving the 
desired gauge hierarchy). The contributions coming from first term of
(\ref{d6nu1}) could be adequately suppressed due to CKM matrix elements
[note that first term in (\ref{d6nu1}) contains extra multiplier
$(V^+)_{n1}$]. In sect. 3, for $SU(5)$ GUT example we will precisely show
this. 
Due to same reasons the contributions from terms of 
(\ref{d6e1}) are suppressed. Since these terms would induce $p\to K\mu $
decays, the
additional inhensiment factor (of the order of 10) in proton lifetime
come from the hadronic
matrix element, which correspond to proton decay with emission of charged
lepton.

As far as the right handed  $d=5$ operators 
${u^c}^i{u^c}^j{d^c}^m{e^c}^n$ are
concerned,
the dominant contribution from them comes through higgsino
dressings. 
Due to antisymmetry in respect of color in this $d=5$ operator, $u^c$
states should be taken from the different generations. So, either $c^c$ or
$t^c$ will appear. Thay must not be appeared in an external line after
dressing, otherwise the diagram would not be relevant for nucleon decay.
Since first two generation sparticles and also all 
${\tl{d^c}}^i$ states 
are in $10$~TeV range, relevant diagram would be that one which inside the
loop contain $\tl{t}^c$ and $\tl{\tau }^c$ states (the latter is
neccessarily light within $SU(5)$ GUT, because it comes from $10_3$ plet).
Due to suppressed mixings of third generation states with first and second
generations and also due to small Yuakawas (in small $\tan \bt $ regime)
appearing due to higgsino dressing, the suppression of nucleon decay can
be guaranteed also for this case [this is shown in sect. 3 on an $SU(5)$
example].

In general, within this scenario universality does not hold and
nucleon decays through gluino dressings would occur. However, 
heavy sparticles will still play a crucial role in suppression of nucleon
decay. Which contribution is dominant depends on the details of
the scenario and in sect. 3 we study this issue
within a realistic $SU(5)$ model.

Thus, thanks to the anomalous ${\cal U}(1)$
symmetry, in addition to avoiding dangerous FCNC, one can also obtain 
adequate
suppression of nucleon decay. Interestingly, this disfavors the large 
$\tan \beta $ regime which could be a characteristic feature of this 
class of models.

\subsection{Possible neutrino oscillation scenarios}

As we previously mentioned, MSSM does not fix the values of $n, k, p$ in
(\ref{charges}), (\ref{qch}). Because of this, the solar neutrino oscillation
scenario is not
specified. According to (\ref{solosc}) both 
LMA and LOW are possible solutions. Namely, for
$n=3$ we have $\De m_{12}^2\sim 10^{-5}~{\rm eV}^2$, which corresponds to
LMA. $n=6$ gives $\De m_{12}^2\sim 10^{-7}~{\rm eV}^2$, which is the
scale
for the LOW solution

In SUSY $SU(5)$ GUT, due to unified $10$, $\bar 5$ multiplets:
\beq
Q_q=Q_{e^c}=Q_{u^c}=Q_{10}~,~~~~~
Q_l=Q_{d^c}=Q_{\bar 5}~.
\label{chsu5}
\eeq
Hierarchies of the CKM matrix elements in (\ref{ckm}) dictate the
relative ${\cal U}(1)$ charges of the $10$-plets
\beq
Q_{10_3}=0~,~~~Q_{10_2}=2~,~~~Q_{10_1}=3~,
\label{ch10}
\eeq
while the Yukawa hierarchies (\ref{ulam})-(\ref{elam}), together with
(\ref{ch10}), require that
\beq
Q_{\bar 5_3}=Q_{\bar 5_2}=k~,~~~~Q_{\bar 5_1}=k+2~.
\label{ch5}
\eeq
Comparing (\ref{chsu5})-(\ref{ch5}) with (\ref{charges}),
(\ref{qch}) we see that the minimal $SU(5)$ GUT fixes $n$ and $p$
as
\beq
n=2~,~~~~~p=0~.
\label{nk}
\eeq
The mass squared splitting in 
(\ref{solosc}) then equals 
$\De m_{12}^2\sim 10^{-4} {\rm eV}^2$, which is a reasonable scale for 
LMA scenario. 
Therefore, realisation of our bi-maximal mixing scenario in
the framework of $SU(5)$ GUT dictates that the LMA scenario is responsible
for
the solar neutrino deficit (more detailed discussion of $SU(5)$ GUT will 
be presented in the next section). 
The same conclusion can be reached for 
$SO(10)$ GUT where we have three $16$-plets of chiral 
supermultiplets which unify the quark-lepton superfields. We do not present the
details here but refer the reader to \cite{6}, where an
explicit $SO(10)$ model with anomalous ${\cal U}(1)$
flavor symmetry is considered  for explanations of fermion masses, their mixings,
as well as neutrino anomalies.

For  models in which an anomalous flavor 
${\cal U}(1)$ also provides SUSY breaking  soft masses, from
(\ref{masssc}) we should require $Q_{\tl{q}, \tl{l}}>0$. On the
other hand, 
from (\ref{solosc}), the realization of LMA and LOW respectively
require $n=3$ and $n=6$ (if $p\stackrel{>}{_{-}}1$), both of which
 guarantee $Q_{e^c}\stackrel{>}{_{-}}0$. 
Therefore, a scenario in which an anomalous
flavor ${\cal U}(1)$ mediates SUSY breaking permits  LMA and LOW  
oscillations for solar neutrinos (the large angle vacuum oscillation
would not be realized within this scenario, since it requires $n=10$
giving negative $Q_{e^c}$).

\section{An $SU(5)$ Example}

Let us now consider in detail a SUSY $SU(5)$ GUT and
show how things discussed in the previous section work out
in practice. We also present here the possibility for avoiding
the problematic
asymptotic mass relations $\hat{M}^0_d=\hat{M}^0_e$ (for first and
second families).

The three families of matter in $(10+\bar 5)$ representations 
have the transformation properties as given in (\ref{ch10}), 
(\ref{ch5}),
while the scalar superfields $\bar H(\bar 5)+H(5)$  have 
$Q_{\bar H}=Q_H=0$.
The couplings relevant for the generation of up, down quark and charged
lepton masses respectively are  given by
\begin{equation}
\begin{array}{ccc}
&  {\begin{array}{ccc}
\hspace{-5mm}~~10_1 & \,\,10_2  & \,\,10_3

\end{array}}\\ \vspace{2mm}
\begin{array}{c}
10_1 \\ 10_2 \\10_3
 \end{array}\!\!\!\!\! &{\left(\begin{array}{ccc}
\,\,\epsilon^6~~ &\,\,\epsilon^5~~ &
\,\,\epsilon^3
\\
\,\,\epsilon^5~~   &\,\,\epsilon^4~~  &
\,\,\epsilon^2
 \\
\,\,\epsilon^3~~ &\,\,\epsilon^2~~ &\,\,1
\end{array}\right)H }~,
\end{array}  \!\!  ~~~~~
\label{upsu5}
\end{equation}
\begin{equation}
\begin{array}{ccc}
 & {\begin{array}{ccc}
\hspace{-5mm}\bar 5_1~ & \,\,\bar 5_2 ~~ & \,\,\bar 5_3 ~~ 

\end{array}}\\ \vspace{2mm}
\begin{array}{c}
10_1 \\ 10_2 \\10_3
 \end{array}\!\!\!\!\! &{\left(\begin{array}{ccc}
\,\,\epsilon^5~~ &\,\,\epsilon^3~~ &
\,\,\epsilon^3
\\
\,\,\epsilon^4~~   &\,\,\epsilon^2~~  &
\,\,\epsilon^2
 \\
\,\,\epsilon^2~~ &\,\,1~~ &\,\,1
\end{array}\right)\epsilon^k \bar H }~.
\end{array}  \!\!  ~~~~~
\label{downsu5}
\end{equation}
Upon diagonalization of (\ref{upsu5}), (\ref{downsu5}) we obtain
the desirable hierarchies (\ref{ulam})-(\ref{ckm}) and 
$\lam_b\sim\lam_{\tau }\sim \ep^k$.

The reader will note, however,  that
(\ref{downsu5}) implies the asymptotic mass relations 
$\hat{M}_d^0=\hat{M}_e^0$, which are unacceptable for the two 
light families.
This is readily avoided through the mechanism suggested in 
\cite{9}, \cite{8} by employing two pairs of 
$(\overline{15}+15)_{1,2}$ matter states. Namely, with 
${\cal U}(1)$ charges
\beq
Q_{15_1}=-Q_{\overline{15}_1}=3~,~~~~~~
Q_{15_2}=-Q_{\overline{15}_2}=2~,
\label{ch15}
\eeq
consider the couplings 
\begin{equation}
\begin{array}{cc}
 & {\begin{array}{ccc}
10_1&\,\,10_2&\,\,10_3~~~
\end{array}}\\ \vspace{2mm}
\begin{array}{c}
\overline{15}_1\\ \overline{15}_2

\end{array}\!\!\!\!\! &{\left(\begin{array}{ccc}
\,\, 1~~&
\,\,  0~~ &\,\, 0
\\
\,\, \epsilon ~~ &\,\,1~~&\,\, 0~
\end{array}\right)\Sigma }~,
\end{array}  \!\!~
\begin{array}{cc}
& {\begin{array}{cc}
15_1&\,\,
15_2~~~~~
\end{array}}\\ \vspace{2mm}
\begin{array}{c}
\overline{15}_1 \\ \overline{15}_2

\end{array}\!\!\!\!\! &{\left(\begin{array}{ccc}
\,\, 1~~
 &\,\,0
\\
\,\, \epsilon~~
&\,\,1
\end{array}\right)M_{15}~,
}
\end{array}
\label{1015}
\end{equation}                
where $\Sigma $ is the scalar $24$-plet whose VEV breaks $SU(5)$ 
down to $SU(3)_c\times SU(2)_L\times U(1)_Y$.
For $M_{15}\sim \langle \Sigma \rangle $, we see that the `light'
$q_{1,2}$ states reside both in $10_{1,2}$ and $15_{1,2}$
states with similar `weights'. At the same time, the other light states 
from $10$-plets ($u^c$ and $e^c$) will not be affected because the 
$15$-plets do not contain fragments with the relevant quantum numbers. Thus,
the relations $m_s^0=m_{\mu }^0$ and $m_d^0=m_e^0$ are avoided, while 
$m_b^0=m_{\tau }^0$ still holds since the terms in (\ref{1015})
do not affect $10_3$.

As far as the sparticle spectrum is concerned, since the superfields
$10_3, \bar H, H$ have zero ${\cal U}(1)$ charges, the soft masses of 
their scalar components will be in the $1$~TeV range,
\beq
m_{\tilde{10}_3}\sim m_{\bar H}\sim m_{H}
\sim m_{3/2}=1~{\rm TeV}~,
\label{soft3h}
\eeq
while for $10_{1,2}$ and $\bar 5_1$ we have
\beq
m_{\tilde{10}_1}\sim m_{\tilde{10}_2}\sim m_{\tilde{\bar 5}_1}
\sim m\sim 10~{\rm TeV}~.
\label{soft12}
\eeq
The soft masses of the scalar fragments from $\bar 5_{2,3}$
depend on the value of $k$, and for $k\neq 0$, they also will be in the
$10$~TeV range, which is preferred for proton stability.
As far as neutrino oscillations are concerned, as already pointed out in
sect. 1.3, due to $SU(5)$ and ${\cal U}(1)$ charge prescriptions, the
LMA solution is preferred with the texture in (\ref{u1nu}).

\subsection{Nucleon Decay in $SU(5)$}

Turning to the issue of nucleon decay in $SU(5)$ , we 
will take $k\neq 0$ in (\ref{ch5}), which provides soft masses for 
$\bar 5_{2,3}$ states in the $10$~TeV range. Let us first make sure that
this ensures proton stability. As pointed out in sect. 2, diagrams with
slepton inside the loop [$2^{\rm nd}$ term of (\ref{d6nu1}) 
and $1^{\rm st}$ 
term of (\ref{d6e1})] are appropriately suppressed. 

The diagrams
corresponding to the first terms in (\ref{d6nu1}) induce nucleon
decay
with the dominant mode $p\to K\nu_{\mu ,\tau }$ with
$\tl{t}$ and $\tl{b}$ running in the loop.
For our scenario, taking into account the couplings
(\ref{upsu5}), (\ref{downsu5}), (\ref{1015}), the dominant
contribution
accompanying $I$ (from first term in (\ref{d6nu1})) is of order
order of $\ep^{k+2}\lam_tV_{td}V_{ub}\sim \lam_s\lam_tV_{td}V_{ub}$. The
corresponding partial lifetime in units of
$\tau_0\equiv \tau[{\rm min.}~{\rm SUSY}~SU(5)]$ will be
$\tau(p\to K\nu_{\mu ,\tau })=\tau_0\l \fr{\lam_c\sin^2 \te_c}
{\lam_tV_{td}V_{ub}}\r^2$. For the CKM matrix elements we have
\beq 
V_{td}=0.004 - 0.014~,~~~~V_{ub}=0.0025 - 0.0048~,
\la{Vtd}
\eeq
and for their central values we have 
$\tau (p\to K\nu_{\mu ,\tau })=217 \tau_0$, which presumably is
still compatible with the available limits
(of course, for lower values in (\ref{Vtd}) we can have much
longer lived nucleon $\tau (p\to K\nu_{\mu ,\tau })
\sim 2\cdot 10^3\tau_0$).

The second term in (\ref{d6e1}) induces the decay $p\to K\mu $.
It can be verified that the suppression discussed above occurs in this
case too. Furthermore,
additional inhancement of the partial lifetime by a factor of  order
$10$ occurs from the hadronic matrix element corresponding to proton
decay
with emission of charged lepton.

As far as the right handed $u^cu^cd^ce^c$ type $d=5$ operators are
concerned, due to arguments presented at the end of sect. 2, the relevant
$d=5$ operator will be $\tl{\tau }^c\tl{t}^cu^cs^c$ which in our $SU(5)$
model, according to (\ref{upsu5}), (\ref{downsu5}), (\ref{1015}) will
appear as
$\fr{1}{M_T}\lam_t\lam_b\ep^3\cdot \tl{\tau }^c\tl{t}^cu^cs^c$. After
higgsino
dressing relevant $d=6$ operator will have the form
\beq
\fr{\kappa }{M_T}\lam_t^2\lam_b^2\ep^3V_{td}\cdot 
(\ov{u^c}~ \ov{s^c})(d\nu_{\mu ,\tau })~.
\la{higsdres}
\eeq
$\kappa $ is factor which coincides with those appearing in minimal SUSY
$SU(5)$ (we assume that $M_{\tl{W}}\simeq \mu$-term). Operator
(\ref{higsdres}) leads to $p\to K\nu_{\mu ,\tau }$ and corresponding
partial lifetime is
$\tau (p\to \nu_{\mu ,\tau })=\l \fr{g_2^2\lam_s\lam_c\sin^2 \te_c}
{\lam_b^2\lam_t^2\ep^3V_{td}}\r^2\tau_0$, which for $\lam_b\sim 10^{-2}$
($\tan \bt \sim 1$) and $V_{td}=0.004$ gives 
$\tau (p\to K\nu_{\mu ,\tau })\sim 400 \tau_0$, compatible with
experimental data.

In considered $SU(5)$ GUT due to different ${\cal U}(1)$ charges 
the universality of soft ${\rm mass}^2$ is
violated. Because of this nucleon decay through gluino dressings will not
be canceled (as is the case in minimal $N=1$ SUGRA). So, these diagrams
are important since the proton decay amplitude will be increased by the
factor $\al_3/\al_2$. For our $SU(5)$ scenario the dominant contributions
come from diagrams inside which run the third generation sparticles
(namely $\tl{t}$, $\tl{b}$ from $q_3$). The relevant $d=5$ operators,
obtained after integration of colored triplet higgs fields, will be
\beq
\fr{1}{M_T}\lam_t\lam_b\ep^3\cdot \tl{t}\tl{b}d\nu_{\mu, \tau}~,~~~
\fr{1}{M_T}\lam_t\lam_b\ep^2\cdot \tl{t}\tl{s}d\nu_{\mu, \tau}~,~
\la{op1}
\eeq
\beq
\fr{1}{M_T}\lam_t\lam_b\ep \cdot \tl{t}\tl{b}u \mu~,
\la{op2}
\eeq
which, after gluino dressings, lead to the following four-fermion
operators:
\beq
\fr{\kappa }{M_T}\al_3\lam_t\lam_b\ep^2L^u_{13}
\left [\ep L^d_{23}\cdot \l d \nu_{\mu, \tau }\r \l u s \r +
L^d_{13}\cdot \l s \nu_{\mu, \tau }\r \l u d\r \right ]~,
\la{4op1}
\eeq
\beq
\fr{\kappa }{M_T}\al_3\lam_t\lam_{\mu }\ep L^u_{13}L^d_{23}\cdot
\l u \mu \r \l u s \r~.
\la{4op2}
\eeq
$L^{u, d}$ are unitary matrices that rotate the left-handed up
and down quark flavor states. Operators (\ref{4op1}) and
(\ref{4op2}) respectively lead to the decays $p\to K\nu_{\mu, \tau }$,
$p\to K\mu $, and the corresponding lifetimes are
\beq
\tau (p\to K\nu_{\mu, \tau })=\left [
\fr{\al_2}{\al_3}\fr{\lam_s\lam_c\sin^2 \te_c}
{\lam_b\lam_t\ep^3L^u_{13}L^d_{23}}\right ]^2\tau_0~,
\la{tau1}
\eeq
\beq
\tau (p\to K\mu )=10\left [
\fr{\al_2}{\al_3}\fr{\lam_s\lam_c\sin^2 \te_c}
{\lam_{\mu }\lam_t\ep L^u_{13}L^d_{23}}\right ]^2\tau_0~.
\la{tau2}
\eeq
{}For $V_{cb}$ we have
\beq
V_{cb}=0.036 - 0.046~,
\la{ckmel}
\eeq
which together with (\ref{Vtd}) can dictate that the values of $L^{u, d}$
can vary,
\beq
L^u_{12}~,~L^d_{13}=(1.8 - 3.5)\cdot 10^{-3}~,~~~~
L^d_{23}=(2 - 4)\cdot 10^{-2}~.
\la{ckmvar}
\eeq
Taking the lowest values from (\ref{ckmvar}), the nucleon lifetimes are
\beq
\tau (p\to K\nu_{\mu, \tau })\sim 177\tau_0~,~~~~
\tau (p\to K\mu )\sim 2\cdot 10^3\tau_0~,
\la{lifeT}
\eeq
which are still compatible with the experimental bounds\footnote{The
lifetimes  
can be further increased if say $L^d_{23}$ is more
suppressed
and in the appropriate entry of CKM matrix the main contribution comes
from $L^u$.}, with
the dominant decay mode being $p\to K\nu_{\mu, \tau }$. 

Before concluding, let us note that the Planck scale suppressed baryon 
number 
violating $d=5$ operator $\frac{1}{M_P}q_1q_1q_2l_{2,3}$, which could 
cause unacceptably fast proton decay, is also suppressed, 
since it emerges from the coupling
\beq
\frac{1}{M_P}\left(\frac{X}{M_P}\right)^{8+k}
10_1 10_1 10_2 \bar 5_{2,3}~,
\label{planckd5}
\eeq
with the suppression guaranteed by the ${\cal U}(1)$ symmetry.

\section{Conclusions}

 It is quite remarkable that the introduction
 of an anomalous ${\cal U}(1)$ symmetry within a supersymmetric
 setting can have several far reaching consequences.
 In particular, the flavor and dimension five proton decay
 problems encountered in SUSY models can be overcome. The
 atmospheric and solar neutrino puzzles can be nicely explained 
 via neutrino oscillations. Although we have emphasized the $SU(5)$
 model here, the discussion can be extended to other realistic
 models such as those based on $SO(10)$.

\vs{0.5cm}

\hs{-6.5mm}{\bf Acknowledgements}

\vs{0.1cm} 

\hs{-6.5mm}We are very grateful to the organizers of the NOON2001 
workshop, especially
Professor M. Bando, for creating a very stimulating environment at the
conference and for their generous hospitality. We also thank the
organizers of the post-NOON workshop in Kyoto.

\bibliographystyle{unsrt}

\end{document}